\documentclass[9pt,twocolumn,twoside]{article}
\usepackage{graphicx,amsmath}
\usepackage{graphicx}
\usepackage{dcolumn}
\usepackage{bm}
\usepackage{url}
\usepackage{color}
\usepackage{floatflt}
\usepackage{hyperref}
\usepackage{amssymb,amsmath}
\usepackage{graphicx}
\usepackage{graphics}
\usepackage{wrapfig}
\usepackage{subfigure}
\usepackage{float}
\usepackage{mathabx}
\usepackage{wasysym}
\usepackage[numbers]{natbib}

\definecolor{astral}{RGB}{46,116,181}
\definecolor{dred}{RGB}{102,0,0}

\newcommand\blfootnote[1]{%
  \begingroup
  \renewcommand\thefootnote{}\footnote{#1}%
  \addtocounter{footnote}{-1}%
  \endgroup
}

\begin{document}
\title{A precise relationship among Buller's drop, ballistospore and gill morphology enables maximal packing of
 spores within gilled mushrooms.}
\author{Martina Iapichino$^1$, Yen Wen Wang$^2$, Savannah Gentry$^2$, Anne Pringle$^2$, Agnese Seminara$^1$}
\maketitle

\blfootnote{$^1$CNRS and Universit\'e C\^ote d'Azur, Institut de Physique de Nice, UMR7010, Parc Valrose 06108, Nice, France}
\blfootnote{$^2$Departments of Botany and Bacteriology, University of Wisconsin-Madison, Madison, WI, USA}
\blfootnote{M.I., A.P and A.S. designed research;  M.I., Y.W.W., S.G. and A.P. performed experiments; M.I. and A.S. performed theoretical analysis and data analysis;  M.I., A.P. and A.S. wrote manuscript.}
\blfootnote{The authors declare no conflict of interest.}
\blfootnote{To whom correspondence should be addressed. E-mail: agnese.seminara@unice.fr}


\noindent{\bf \large Abstract.} {\small Basidiomycete fungi eject spores using a surface tension catapult; a fluid drop forms at the base of each spore and after reaching a critical size, coalesces with the spore and launches it from the gill surface. Although basidiomycetes function within ecosystems as both devastating pathogens and mutualists critical to plant growth, an incomplete understanding of ballistospory hinders predictions of spore dispersal and impedes disease forecasting and conservation strategies. Building on a nascent understanding of the physics underpinning the surface tension catapult, we first use the principle of energy conservation to identify ejection velocities resulting from a range of Buller's drop and spore sizes. 
We next model a spore's trajectory away from a basidium and identify a specific relationship among intergill distances and Buller's drop and spore radii enabling the maximum number of spores to be packaged within a minimum amount of gill tissue. We collect data of spore and gill morphology in wild mushrooms and we find that real species lie in a region where, in order to pack the maximum number of spores with minimum amount of biomass, the volume of Buller's drop should scale as the volume of the spore, and its linear size should be about half of spore size. Previously published data of Buller's drop and spore size confirm this finding. Our results suggest that the radius of Buller's drop is tightly regulated to enable maximum packing of spores. 
}\\

\noindent{\bf \large Introduction.} Molds, yeasts and mushrooms are ubiquitous across Earth. Estimates of the number of fungal species range from 500,000 to almost 10 million \cite{hawk2017} and fungi in ecosystems function as decomposers, mutualists and pathogens. Emerging fungal-like diseases endanger crops as well as wild plants and animals, threatening food security, but also altering forest dynamics and contributing to the extinction of animals. Losses cost millions of dollars of damage \cite{pennisi2010,Kupferschmidt_2012,Fisher_2012}.

\begin{figure}[h!]
\begin{center}
{\includegraphics[width=0.5\textwidth]{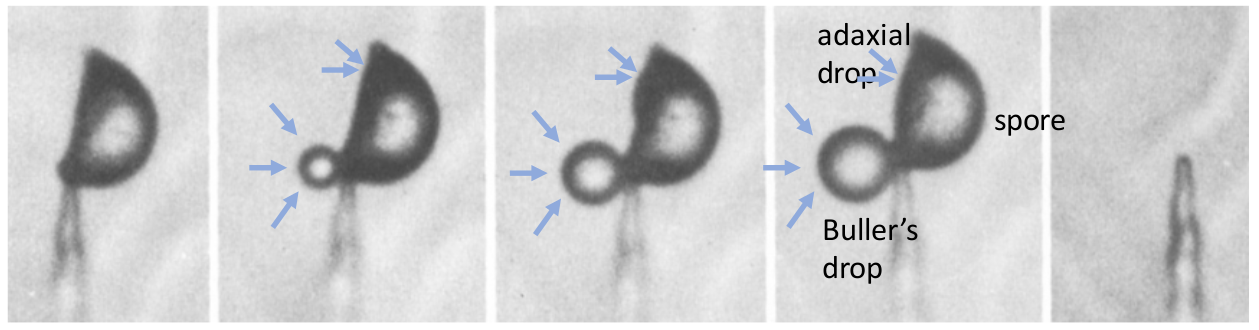}}
\caption{\footnotesize Diagram illustrating our current understanding of the surface tension catapult. Over short time scales, Buller's drop forms and grows at the base of the spore. At a critical size, the drop collapses onto the spore end reduces its surface, thus releasing energy. The released energy is converted into kinetic energy of the spore. The size of Buller's drop, together with material parameters, determines the speed of spore discharge. Image adapted from \cite{Websteretal1984}.} 
\label{fig:coalescence}
\end{center}
\end{figure}

Fungal bodies are immobile, typically hidden within substrates. Fungi use spores to reproduce and travel away from a particular habitat. Spores are carried in currents more or less far away from a source and when a spore lands in a favorable environment, it germinates and begins or extends the life cycle. Basidiomycetes are among the most common fungi, encompassing \emph{Puccinia graminis} and \emph{Phakopsora pachyrhizi} (Asian soybean rust). The phylum is defined by the basidiospore. Typically, basidiospores are launched via a surface tension catapult. In species with mushrooms, spores grow in groups of four from basidia arranged along the gills or pores of a mushroom cap's underside, each spore attached to a sterigma. A drop of liquid, known as Buller's drop, forms extracellularly at the base of each spore by condensation of water. Buller's drop then collapses onto another adaxial drop formed on the longitudinal axis of the spore itself (see sketch in Figure~\ref{fig:coalescence}). Upon coalescence, surface energy is converted into kinetic energy and transmitted to the spore which is ejected horizontally away from the basidium and sterigma. The spore decelerates to rest in few milliseconds and then falls vertically between two gills or within the pore.

\noindent Ballistospore discharge is a phenomenon that has fascinated scientists for over a century. It was first observed by Schmitz \cite{schmitz1843}. In the 20th century, Buller described the phenomenon in more detail, observing the development and discharge of the spore, describing the secretion of the drop at the hilum (the junction between the spore and its sterigma) and the subsequent discharge of the spore together with the drop \cite{buller}. This drop is now referred to as Buller's drop and the discharge understood as a ``surface tension catapult''. Progress in understanding of the anatomy and the physics of the surface tension catapult was enabled by the development of cameras. Webster et al.~provided photographic evidence of Buller's drop forming at the hilar appendix just before discharge and proposed a two-phase mechanism for spore ejection: the first phase involving Buller's drop enveloping the spore surface, acquiring momentum; the second involving the sharing of momentum and movement of the center of mass of the spore-drop complex due to the rapid wetting \cite{Websteretal1984}. Subsequent works modeled the conversion of surface energy into kinetic energy  with different degrees of complexity and monitored this process with progressively faster cameras \cite{pringle2005,Noblinetal2009,aerodynamics_ballisto,Stolze-Rybczynski2009,liu2017}. Pringle et al.~\cite{pringle2005} 
propose that coalescence occurs between Buller's drop and a second drop present on the side of a spore (adaxial drop), while Noblin et al.~\cite{Noblinetal2009} describe the process through four-stages and estimate that about half of the total surface energy is dissipated overall.
Liu et al recently moved beyond energy balance including simulations of the fluid dynamics within Buller's drop during coalescence as well as experiments with bio-mimetic drops \cite{liu2017}. They found that coalescence occurs in a regime where viscous dissipation in the drop is negligible. Hence energy is not dissipated to set Buller's drop in motion, but presumably to break the hilum apart.
Additionally, they find that pinning of the contact line provides directionality of spore/drop complex ejection away from the originating gill. 

It has long been hypothesized that mushrooms form gills to increase the surface area for spore production and pack the maximum number of spores with minimum biomass investment \cite{buller,mcknight1991,gills}. To achieve this optimal morphology, the size of Buller's drop, the size of the spore and the distance between gills must be finely coordinated. While spore size and gill distance may be under genetic control \cite{kuesliu2000}, Buller's drop forms extracellularly \cite{webster1989}. Whether and how fungi control Buller's drop size remains unknown, although data reporting characteristic sizes of Buller's drop for different species suggest individual species do control size \cite{aerodynamics_ballisto,thesis_jessica,pringle2005,Stolze-Rybczynski2009}. 

Here we recapitulate the theory that relates the ejection velocity and flight time with the horizontal distance traveled by the spore at the moment of the ejection before falling between two gills \cite{buller,pringle2005,Noblinetal2009,aerodynamics_ballisto,Stolze-Rybczynski2009,liu2017}. We combine the expressions for ejection speed and flight time, to highlight their dependence on the sizes and densities of the spore and of Buller's drop. We first identify the dependence of ejection velocity on Buller's drop radius. We then use this expression to pinpoint the criteria corresponding to maximum spore packing in the phase space composed of spore radius, intergills distance and Buller's drop radius. To compare models to data, mushrooms of eight different species are collected and the statistics of spore size and gill distance for each specimen is measured. By comparing real morphologies to the prediction generated from our theory we discover that all species lie in a region of the parameter space for which the radius of Buller's drop that maximizes spore packing is about 55\% with respect to spore radius. Previously published data suggest that for at least 13 species monitored directly, Buller's drop scales as $\sim 32\% $ spore size, close to our prediction for the Buller's drop that ensures maximum packing.
This work connects the morphologies of spores, Buller's drops and gills, and opens new insights into the principles that shape  ballistospore discharge while opening up the question of how these constraints are implemented in practice.

\section*{Results}

\begin{figure}
\includegraphics[width=0.5\textwidth]{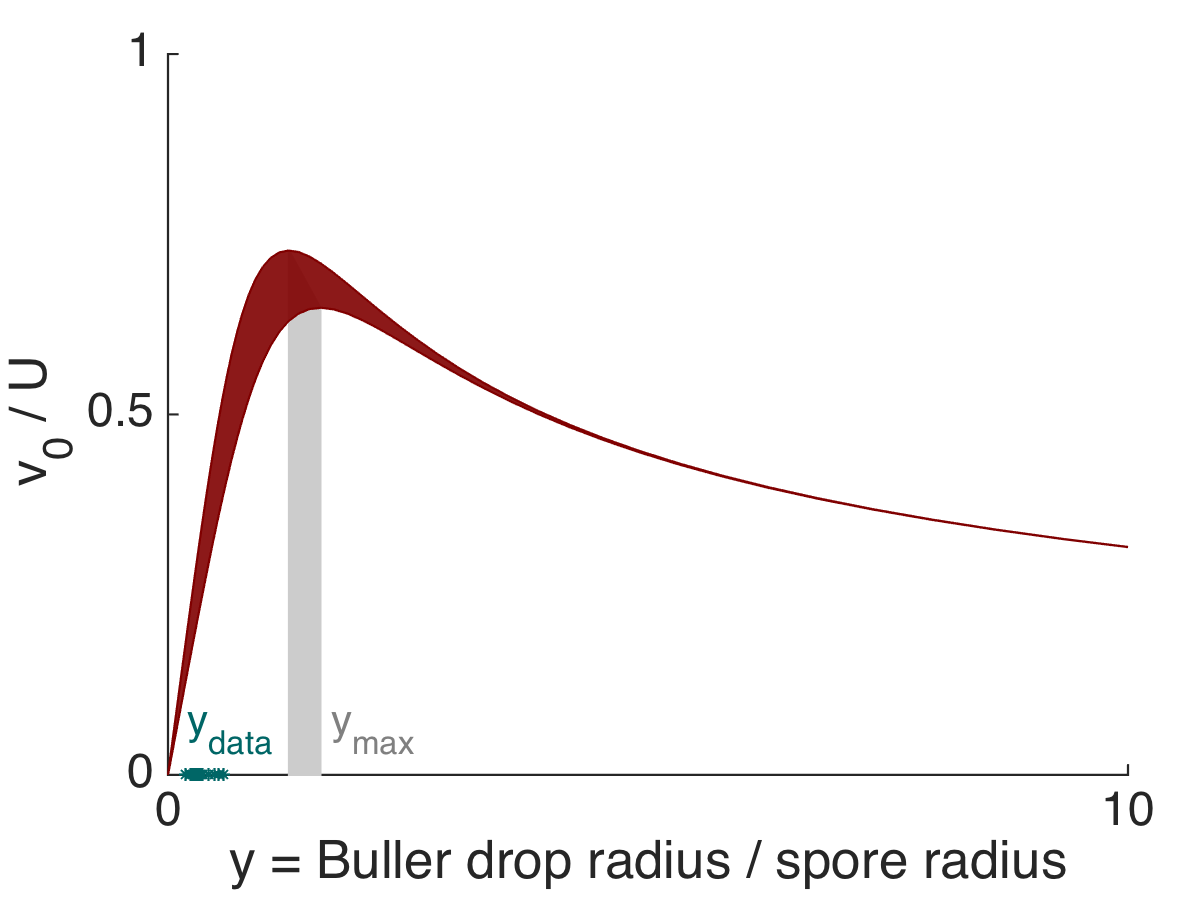}
\caption{\footnotesize Energy balance from eq~\eqref{eq:v0} predicts discharge speed $v_0$ as a function of $y$ defined as the ratio of Buller's drop radius $R_B$ divided by spore equivalent radius $R_s$. Velocity peaks at $y_{\text{max}} = (2 \beta)^{1/3} = 1.26$ to $1.59$ for $\beta$ ranging from 1 to $2$, where $\beta$ is the ratio of spore to drop density. 
The same ejection speed is attained for two values of $y$, one on either side of the maximum. Experimental data of $y$ all lie to the left of the peak, suggesting evolution has favored smaller drops. 
} 
\label{fig:velocity}
\end{figure}

\noindent{\bf Ejection speed.} The surface tension catapult realizes maximum ejection speed when the spore and Buller's drop have nearly the same volume. The two drops that coalesce to power the surface tension catapult are made of condensed water vapor and form after secretion of hygroscopic substances by the spore. 
When Buller's drop coalesces with the adaxial drop, the resulting reduction of surface area provides the surface energy to accelerate the spore. Because the adaxial drop is pinned to the surface of the spore, Buller's drop accelerates towards the distal tip of the spore. Once the coalesced drops reach the tip of the spore, capillarity and contact line pinning decelerate water and its momentum is transferred to the spore. Momentum transfer causes the force that breaks the hilum and results in spore ejection away from the basidium.  
The release of surface energy by coalescence is  $\sim\pi\gamma R_B^2$, where $\gamma$ is surface tension, $R_B$ is Buller's drop radius. 
By balancing surface energy to kinetic energy of the spore - drop complex, we obtain:
\begin{equation}
v_0 = 
U \sqrt{\dfrac{y^2}{y^3+\beta}}
\label{eq:v0}
\end{equation}
where $v_0$ is the ejection velocity, $U= \sqrt{3\alpha\gamma/(2\rho_B R_s)}$; $y=R_B/R_s$, $R_B$ is Buller's drop radius and $R_s$ is the radius of a sphere with the same volume as the spore; $\beta = \rho_B/\rho_s$; $\rho_B$ and $\rho_s$ are densities of Buller's drop and spore respectively. The parameter $\alpha$ signifies that a fraction of available energy is dissipated in the process of breaking the spore apart from the hilum, the structure that holds it attached to the gill. We will consider $\alpha=0.23$, which is the average among the values of efficiency previously measured \cite{thesis_jessica}. Viscous dissipation during the dynamics of coalescence can be neglected because ballistospory operates in a regime of low Onhesorge number \cite{liu2017}. We realized that the simple energy balance discussed at length in the literature and recapitulated in equation~\eqref{eq:v0} predicts that there is a radius of Buller's drop that maximizes $v_0$ (see Figure 2). By zeroing the derivative in \eqref{eq:v0} we obtain the size of Buller's drop that maximizes ejection speed:
\begin{equation}
y_{\text{max}}=(2 \beta)^{1/3}
\label{eq:ymax}
\end{equation}
\noindent and considering spores with density once or twice the density of water\cite{hussein}, $\beta=1$ to 2, implying that at $y_{\text{max}}$ Buller's drop radius is comparable to the equivalent radius of the spore $R_B\sim 1.26 R_s$ to $1.59 R_s$(Figure~\ref{fig:velocity}, grey vertical mark labeled $y_{\text{max}}$). 
Note that at $y_{\text{max}}$ the ejection speed is controlled robustly, i.e.~it becomes insensitive to small deviations from the exact value of Buller's drop size. 
Buller's drop is generally assumed to scale with spore length \cite{aerodynamics_ballisto} and this scaling appears to hold for at least 13 species of basidiomycetes as shown in \cite{aerodynamics_ballisto,thesis_jessica,pringle2005,Stolze-Rybczynski2009}. Supplementary Figure 1 shows these published data, as a function of spore equivalent radius $R_s$, pointing to $y_{\text{data}}= R_B/R_s \sim 0.35 \pm 0.11$ where we report average $\pm$ standard deviation.
$y_{\text{data}}$ are represented on the horizontal axis in Figure~\ref{fig:velocity}, suggesting these fungi do not operate at maximum ejection speed, but rather remain on the rising slope preceding the maximum. 
\\

\begin{figure}[h!]
\includegraphics[width=0.5\textwidth]{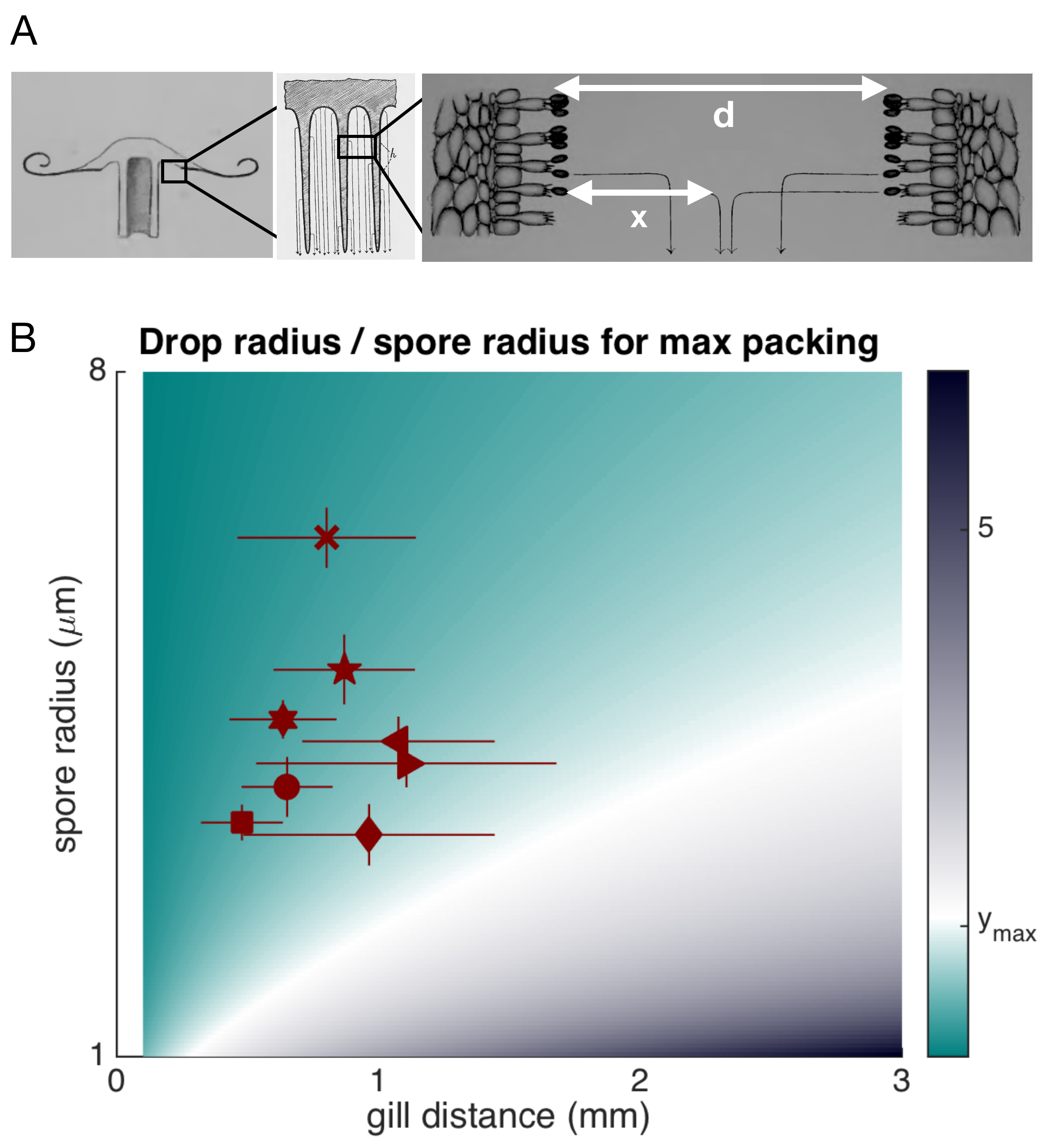}
\caption{\footnotesize Optimal morphology of mushroom caps. (A) Sketch of a cross section of a mushroom, close up of gills and magnified view of adjacent gills with basidia and basidiospores. Several trajectories of individual spores (sporabolas) are represented with black arrows; trajectories traced by Buller in 1909 \cite{buller}. Maximum packing implies that spores initially travel a distance $x = v_0 \tau$ to reach the midpoint between two opposite gills $d =2 v_0 \tau$ with $v_0$ and $\tau$ given by Equations~\eqref{eq:v0} and \eqref{eq:tau}. (B) Prediction for normalized Buller drop radius at maximum packing, $y_{\text{pack}}$, obtained by numerically solving Equation~\eqref{eq:theory} with $\beta=1.2$, $\bar{\beta}=0.001$ and $\alpha=23$\%. $y_{\text{pack}}$ is color coded from 0 (cyan) to $10$ (dark blue), and white marks normalized Buller drop radius at maximum velocity from Equation~\eqref{eq:ymax}. Red symbols correspond to data of intergill distance and spore equivalent radius from 8 species collected and analyzed in the present study (see Figure 4).
The predicted radius of Buller's drop that maximizes packing for the 8 collected species is $y_{\text{pack}} \sim 0.56 \pm 0.20$, which compares well to measured values of Buller's drop size pointing to $y \sim 0.35 \pm 0.11$, where we report average $\pm$ standard deviation.} 
\label{fig:maxpack}
\end{figure}

\noindent{\bf Maximum spore packing.} 
Once the spore-drop complex is ejected, it is soon decelerated by air drag and its relaxation time is well approximated by the Stokes time \cite{stokes,aerodynamics}:
\begin{equation}
\tau 
= T (y^3 + 1)^{2/3}
\label{eq:tau}
\end{equation}
\noindent where we have considered the complex as an equivalent sphere with volume equal to the sum of the spore and drop volumes. Here, $T=2R_s^2 / (9\nu \bar{\beta})$, $\nu$ is the air kinematic viscosity, $\bar{\beta}$ is the density of air divided by the density of the spore-drop complex. 
After discharge, spores travel horizontally a distance $x=v_0 \tau$, with $v_0$ and $\tau$ from equations~\eqref{eq:v0} and \eqref{eq:tau} and then stop abruptly and start to sediment vertically out of the gills, following a trajectory commonly known as ``sporabola'' (represented in Figure~\ref{fig:maxpack}A). In order to successfully escape, spores should first travel horizontally far enough to avoid sticking to the spores and basidia underneath. If $x$ is indeed dictated by this safety criterion, then the distance between two opposite gills, $d$, should be at least twice $x$, hence $d>2x$. To pack as many spores as possible and avoid inefficient empty spaces, 
the distance between gills must be close to this minimum value:
\[
d \sim 2 v_0 \tau
\]
\noindent Plugging in the values of $v_0$ and $\tau$ given by equations~\eqref{eq:v0} and \eqref{eq:tau} we obtain:
\begin{equation}
\frac{d}{2UT} = 
\Bigl(\frac{y_{\text{pack}}^2}{y_{\text{pack}}^3 + \beta} \Bigr)^{1/2} (y_{\text{pack}}^3+1)^{2/3} 
\label{eq:theory}
\end{equation}
For any combination of spore density and radius as well as intergill distance, equation~\eqref{eq:theory} predicts the optimal radius of Buller's drop normalized by spore radius, $y_{\text{pack}}$, that achieves maximum packing. We solve numerically Equation~\eqref{eq:theory} and show the result for $y_{\text{pack}}$ in Figure~\ref{fig:maxpack} for different combinations of intergill distance and spore radius, assuming $\beta=1.2$, $\alpha=0.23$, $\bar{\beta}=10^{-3}$ color-coded from 0 (cyan) to 10 (dark blue). The value of $y_{\text{max}}$ from Equation~\eqref{eq:ymax}  that maximizes ejection speed is marked in white for $\beta=1.2$, for reference. 
\\

\begin{figure}[h!]
\includegraphics[width=0.5\textwidth]{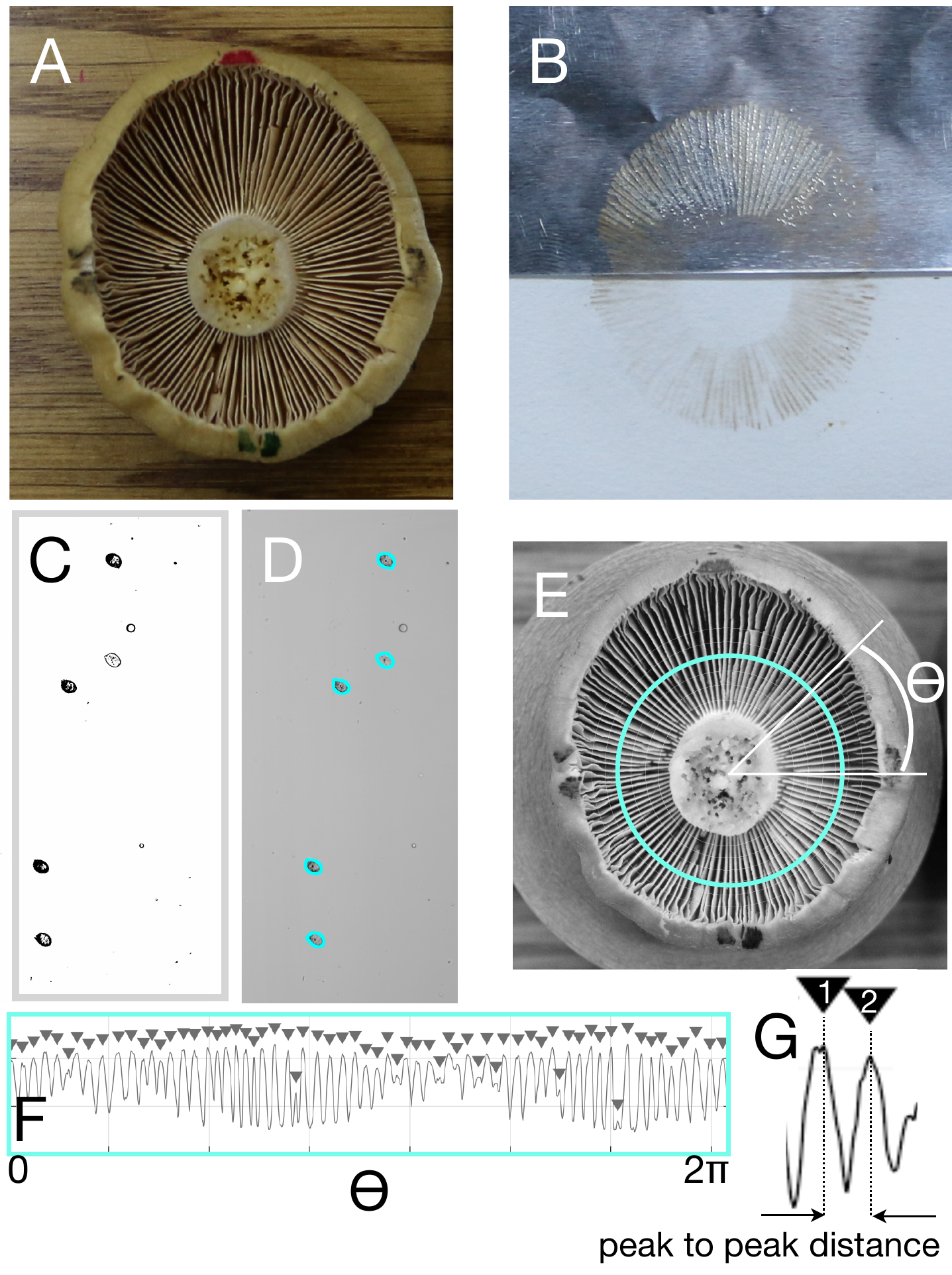}
\caption{\footnotesize Data collection. (A) Picture of wild isolate of mushroom cap. (B) Spore print obtained by deposition of the spore cap on aluminum foil overnight. (C) Confocal microscope image of a sample of spores from the spore print. (D) Segmentation of spore image to recover spore contour. (E) Concentric circle around the center of the cap where gill distance is measured and definition of azimuthal angle $\theta$. (F) Grey scale value from image in panel E, as a function of azimuthal angle $\theta$. (G) Close up image showing locations of two peaks in the grey image, marked automatically by arrows 1 and 2 (above). Gill distance is defined as the distance between peaks minus their  width (see Materials and Methods).} 
\label{fig:collection}
\end{figure}
\begin{figure}
\includegraphics[width=0.5\textwidth]{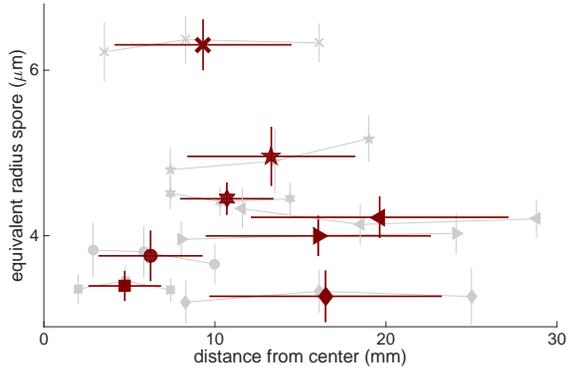}
\caption{\footnotesize Results of data analysis. Spore size does not vary across a single mushroom cap.} 
\label{fig:avspores}
\end{figure}
\begin{figure}
\includegraphics[width=0.5\textwidth]{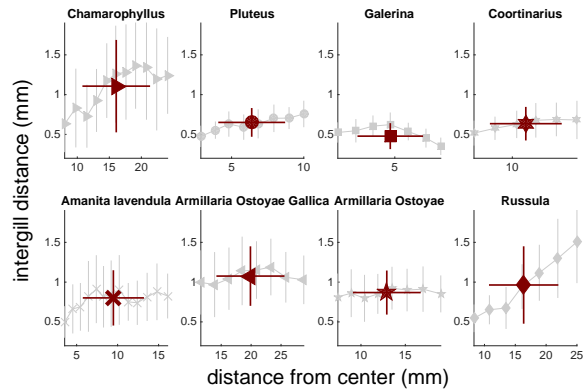}
\caption{\footnotesize Results of data analysis. Average gill spacing varies little with distance from the center of the cap. The only exception is \emph{Russula cremicolor} which has no secondary gills.}
\label{fig:avintergill} 
\end{figure}

\noindent{\bf Data collection and data analysis.} \\
To place real species on the phase space generated by the theory, we collected data of spore and gill morphology for eight wild mushroom isolates. 
We isolate mushroom caps (Figure~\ref{fig:collection}A), let them sit overnight on aluminum foil, resulting in what is called a spore print (Figure~\ref{fig:collection}B) and then isolate samples of spores from different regions of the mushroom. Spores are imaged under confocal microscopy (Figure~\ref{fig:collection}C), and images are analyzed with a standard segmentation postprocessing using imageJ to isolate contours of spores (Figure~\ref{fig:collection}D). Spore area $S$ is computed from these images and radius is obtained from the area $R_s=\sqrt{S/\pi}$. To measure gill distance, we first identify the center of the cap by eye. We then draw several circles, between 6 and 10 depending on the size of the cap, around the center of the cap  (Figure~\ref{fig:collection}E). Grey values along the circles are obtained  (one example in Figure~\ref{fig:collection}F) and the profile of the grey value analyzed to define the distance $d$ between the gills as the peak to peak distance minus the width of the peaks (see close up of two peaks in Figure~\ref{fig:collection}G, and Materials and Methods). \\ 
The collected data show that spore size varies from  species to species, but does not vary across a single mushroom cap, suggesting that mushrooms tend to produce spores of the same size in a single fruit body  (Figure~\ref{fig:avspores}). The average intergill distance varies little with distance from the center of the cap, with the exception of \emph{Russula cremicolor}, which is the only species with no secondary gills in our collection, consistent with previous models and experiments \cite{phillips91,gills}. The intergill distance varies from about 0.25~mm to 1.5~mm (Figure~\ref{fig:avintergill}) with no obvious correlation with the size of the mushroom cap.
We use these data to compute average and standard deviation of spore radius and intergill distance across a single individual. The experimental data are superimposed to the theory for maximum spore packing. 
The 8 species tested in this study fall in a region where, if gill morphology is optimized for maximum spore packing, then Buller's drop radius is $R_B\sim 0.55 R_s$, consistent with previously published data pointing to $R_B\sim 0.33 R_s$ (Figure~\ref{fig:maxpack}B and Figure~\ref{fig:velocity}). 
\\

\noindent{\bf Conclusions.} 
Gilled mushroom have long been hypothesized to have intricate morphologies to maximize the surface to volume ratio and pack the maximum number of spores with minimum amount of biomass. In order to comply with this hypothesis, the horizontal range that spores travel upon ejection must be finely tuned to land spores midway in between two opposite gills. Spore range is dictated by the dimension of Buller's drop and its density relative to the dimension and density of the spore. We find that real species populate a region of the phase space where the radius of Buller's drop that maximizes spore packing achieves velocities that are  smaller than the maximum ejection speed $R_B \sim 0.55 R_s$, while at maximum ejection speed $R_B \sim 1.3$ to $1.6 R_s$.
This conclusion is backed from data previously published in the literature, suggesting that Buller's drop radius does indeed scale with spore dimensions, and is smaller than the value that maximizes ejection speed $R_B \sim 0.32 R_s$.   
Further data monitoring spore, gills and Buller's drop morphologies and densities at the same time are needed to find how close are species to maximum packing.
All data to date are consistent with the hypothesis of maximum spore packing, suggesting that Buller's drop radius is finely tuned to control range and speed. How this fine tuning might function, in a process that is purely extracellular, in the face of fluctuations in the environmental conditions remains a fascinating question for future research.

\begin{table*}
\caption{\footnotesize List of collected species, location that these specimens were collected from, number of spores imaged and analyzed, corresponding symbol used in Figure~\ref{fig:maxpack},\ref{fig:avspores}-\ref{fig:avintergill}.}
\label{tab:species}
\begin{tabular}{p{0.3\textwidth}p{0.38\textwidth}p{0.1\textwidth}p{0.1\textwidth}}
\hline
{\footnotesize \bf Collected species} & {\footnotesize \bf Location} & {\footnotesize \bf \# spores}& {\footnotesize \bf symbol}\\
\hline
{\footnotesize \emph{Camarophyllus borealis}}& {\footnotesize Huron Mountain Club }& 231 &  $\pmb \RHD$\\
{\footnotesize \emph{Cortinarius caperatus}}& {\footnotesize Huron Mountain Club }& 1180 & $\pmb\bigvarstar$\\
{\footnotesize \emph{Amanita lavendula}}&{\footnotesize  Huron Mountain Club }& 155 & $\pmb\times$\\
{\footnotesize \emph{Armillaria mellea sp. complex}}& {\footnotesize Huron Mountain Club }& 301 &$\pmb\bigstar$\\
{\footnotesize \emph{Armillaria mellea sp. complex}}& {\footnotesize Huron Mountain Club }& 257 & { $\pmb\LHD$}\\
{\footnotesize \emph{Mycena sp.}} & {\footnotesize UW-Madison Lakeshore Natural Preserve }& 530 & {\Large$\pmb\bullet$}\\
{\footnotesize \emph{Russula sp.}} & {\footnotesize UW-Madison Lakeshore Natural Preserve }& 1053 & {\Large $\pmb\blackdiamond$}\\
{\footnotesize \emph{Galerina marginata/autumnalis}} & {\footnotesize UW-Madison Lakeshore Natural Preserve }& 1159 &  {\Large$\pmb\sqbullet$}\\
\hline
\end{tabular}
\end{table*}

{\small
\section*{\bf \large Materials and methods.}
\noindent{\bf Data collection and published data}\quad Between the 15th and 17th of September, 2017 we collected mushrooms from lands owned by the Huron Mountain Club, in the Upper Peninsula of Michigan. On the 15th of October, 2017 we collected mushrooms from the University of Wisconsin-Madison Lakeshore Natural Preserve. We collected opportunistically, taking any mushroom that appeared in good shape, but focusing on gilled (not pored) fungi. Unfortunately we were collecting during a particularly dry period, nonetheless, we collected specimens of eight morphologically identified species, listed in Tab.~\ref{tab:species}. 
We integrated our data with data from the literature where spore dimensions and radius of the Buller's drop were precised ~\cite{aerodynamics_ballisto,thesis_jessica,pringle2005,Stolze-Rybczynski2009}. \\

\noindent{\bf Preparing specimens for morphometrics}\quad On the same day mushrooms were collected, caps were separated from stipes using a scalpel and left face down from 8 to 12 hours on a piece of paper covered with aluminum foil in order to create spore prints. Spore prints are generated when spores fall from gills and settle directly underneath. They reflect the morphology of each collected specimen and the location of stipes and patterns of gill spacing are easily visualized. Spore prints were carefully wrapped in wax paper and taken back to the Pringle laboratory at the University of Wisconsin-Madison. To image spores, three small pieces of aluminum foil, each measuring approximately 1mm x 1mm , were cut (i) from close to each stem,(ii) equidistant between the stem and the cap edge and (iii) from near the edge of each cap. Spores were washed off the foil and suspended in a Tween 80 $0.01 \% $ vol solution. 15 $\mu$l of each spore suspension were then spread right after onto a glass slide and spores imaged. Microscope slides were sealed with nail polish in order to avoid evaporation of Tween and consequent movement of spores during the imaging. To measure distance between gills, a photograph of each cap's underside, with a ruler included in the photograph, was taken immediately after spore printing using a Canon EOS400D. After spore printing and photography, collected mushrooms were dried in a mushroom dryer and stored in the Pringle laboratory.\\

\noindent{\bf Identification of species using DNA barcoding}\quad To identify the taxa of sporocarps, we extract DNA with NaOH extraction method modified from Wang et al. (1993) to amplify internal transcribed spacer. Specifically, the tissues of the sporocarps were ground finely with pestle in $40 \mu l$ of 0.5 M NaOH and
centrifuged at 13,000 rpm for 10 min. Five microliters of supernatant was transferred to
$495 \mu l$ of 100 mM Tris-HCl (pH 8) and centrifuged at 13,000 rpm for another min.
To amplify the internal transcribed spacer, 1 µl of extracted DNA was mixed with $1
\mu l$ of $10 \mu M$ ITS1F (5’-CTT GGT CAT TTA GAG GAA GTA A-3’), 1 µl of 10 µM ITS4 (5’-TCC
TCC GCT TAT TGA TAT GC-3’), $12.5 \mu l$ of Econotaq plus green 2x master mix (Lucigen,
Wisconsin), and $9.5 \mu l$ of nuclease-free water. The reaction mixtures were incubated in
$95 \circ C$ for 5 min, followed with 30 rounds of amplifying reaction, including (1)
denaturation under $95 \circ C$ for 30 s, (2) primer annealing under $50 \circ C$ for 30 s and (3)elongation under 72˚C for 60 s. The reaction ends with 7 min of additional elongation
under $72 \circ C$ and pauses at $4 \circ C$. Amplified internal transcribed spacer were cleaned,
Sanger-sequenced by Functional Biosciences (Wisconsin) and deposited on Genbank
database (https://www.ncbi.nlm.nih.gov/).\\

\noindent{\bf Microscopy and image analysis for spore geometry.}\quad Microscope images of spores were taken and recorded at the Newcomb Image Center at the University of Wisconsin-Madison. Spores were imaged either individually or in groups depending whether a particular microscope field of view housed one or more than one spore using Zeiss Elyra LSM 780 and Zeiss LSM 710 confocal microscopes. Spores were not stained as all species collected proved to be autofluorescent. The laser wavelength used to excite autofluorescence was 405 nm. The average area and average radius of spores of each species were then calculated using an image analysis tool implemented in ImageJ v.~1.51. Pixel's dimension in $\mu$m was obtained from the microscope and the image converted to greyscale (8-bit or 16-bit). Thresholding was done using imageJ to then convert greyscale to binary image, highlight all the spores to be counted and measure the area of each spore as shown  in Figure~\ref{fig:collection}C-D. Spores touching other spores were not measured, nor were particles smaller than $2 \mu m^2$. Particles bigger than $ 2 \mu m^2$ were identified either as spores or not-spores by eye. \\

\noindent{\bf Image  analysis for gill distance.}\quad The distance betweeen gills was measured based on the cross section of gills at various distances from the center of the cap. Image analysis with ImageJ v1.51 and then analyzed with a custom made Matlab R2017b script. 
We first used ImageJ v1.51 to open each picture, set pixel length in $mm$ using the image of the ruler and convert images to greyscale (8-bit or 16-bit). The Oval Profile plugin was implemented to obtain the grey scale profile along oval traces, drawn manually around the mushroom cap center. The area of the ovals was measured to calculate the average distance from the cap center which was used to convert the distance between gills from radiants to mm. The greyscale is sampled at 3600 equally spaced points along the oval. 
The grey scale profile obtained from ImageJ was imported into Matlab and analyzed with the function Findpeaks to first identify the center of the gills as the peaks in the greyscale image. Peaks that were closer than 0.3$^\circ$ were discarded as noise. Visual inspection was applied to check that minor peaks did correspond to gills. Additionally, we quantified gill thickness as the width of the peak, defined as the distance where grey value drops half way below peak prominence, which is a measure of peak height. Distance between two gills is defined as the distance between their centers minus the half width of the two gills.  

}

\section*{Aknowledgements} 
This work was supported by the Agence Nationale de la Recherche Investissements d'Avenir UCA$^{\textrm{\sc \sf \tiny JEDI}}$ \#ANR-15-IDEX-01, by CNRS PICS ``2FORECAST'',  by the Thomas Jefferson Fund a program of FACE and by Global Health Institute - University of Wisconsin-Madison. We would like also to thank Houron Mountain Club for its kind hospitality and Sarah Swanson for all her help and discussions about confocal microscopy.

\bibliography{ref_gill}
\bibliographystyle{unsrtnat}

\end{document}